\documentclass[journal=nalefd]{achemso}

\usepackage{graphicx}
\usepackage[ansinew]{inputenc}
\usepackage{amsmath,amsfonts,amssymb}

\author{Jörg Kröger}
\email{joerg.kroeger@tu-ilmenau.de}
\affiliation{Institut de Physique et Chimie des Matériaux de Strasbourg, Université de Strasbourg, F-67000 Strasbourg, France}
\altaffiliation{On sabbatical leave; permanent address: Institut für Physik, Technische Universität Ilmenau, D-98693 Ilmenau, Germany}

\author{Benjamin Doppagne}
\affiliation{Institut de Physique et Chimie des Matériaux de Strasbourg, Université de Strasbourg, F-67000 Strasbourg, France}

\author{Fabrice Scheurer}
\affiliation{Institut de Physique et Chimie des Matériaux de Strasbourg, Université de Strasbourg, F-67000 Strasbourg, France}

\author{Guillaume Schull}
\email{guillaume.schull@ipcms.unistra.fr}
\affiliation{Institut de Physique et Chimie des Matériaux de Strasbourg, Université de Strasbourg, F-67000 Strasbourg, France}

\title{Fano description of single-hydrocarbon fluorescence excited by a scanning tunneling microscope}

\begin{document}

\begin{abstract}
The detection of fluorescence with submolecular resolution enables the exploration of spatially varying photon yields and vibronic properties at the single-molecule level.
By placing individual polycyclic aromatic hydrocarbon molecules into the plasmon cavity formed by the tip of a scanning tunneling microscope and a NaCl-covered Ag(111) surface, molecular light emission spectra are obtained that unravel vibrational progression.
In addition, light spectra unveil a signature of the molecule even when the tunneling current is injected well separated from the molecular emitter. 
This signature exhibits a distance-dependent Fano profile that reflects the subtle interplay between inelastic tunneling electrons, the molecular exciton and localized plasmons in at-distance as well as on-molecule fluorescence.
The presented findings open the path to luminescence of a different class of molecules than investigated before and contribute to the understanding of single-molecule luminescence at surfaces in a unified picture.

\begin{tocentry}
\begin{center}
\includegraphics[width=0.8\textwidth]{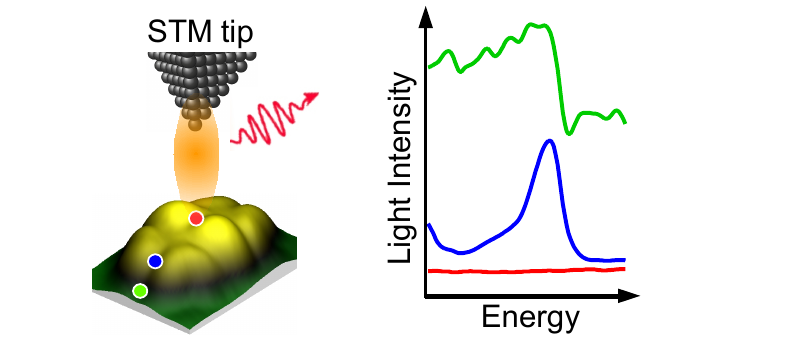}
\end{center}
\label{TOC}
\end{tocentry}
\end{abstract}

Exploring single-molecule luminescence with a scanning tunneling microscope (STM) is an emergent and rapidly evolving research field.
The combination of high spatial resolution in imaging and high energy resolution in optical spectroscopy opens the path to atomic-scale investigations into quantum optics at surfaces.
Since its first observation from molecular crystals \cite{science_262_1425} and from single molecules \cite{science_299_542} many other examples for light emission at the molecular scale followed and are summarized in excellent review articles \cite{ssr_65_129,cr_117_5174}.

In STM-induced light emission (STM-LE) of adsorbed molecules the tip-induced plasmon that forms between the tip and the metal surface \cite{prl_67_3796,zpb_84_269,jjap_31_2465} plays a pivotal role.
As a consequence of the Purcell effect \cite{pr_69_37} it enhances the density of optical modes the molecular excitations may radiatively decay into and, thus, increases the emitted photon intensity.
Therefore, the majority of single-molecule luminescence studies were performed on metal surfaces.
At the same time a seemingly counter-running prerequisite for the observation of the radiative decay of genuine molecular excitations is the efficient reduction of the interaction between the adsorbed molecule and the metal substrate.
It was previously recognized that the contact to a metal surface quenches intramolecular transitions since the fast charge transfer channel to the metal substrate dominates the slow luminescence channel \cite{jpc_88_837}.
Indeed, the STM-LE of molecules adsorbed on metal surfaces reflects the modified radiation of the tip-induced plasmon rather than the sought molecular luminescence \cite{prb_65_212107}.
Therefore, accessing the quantum optical properties of a single molecule requires the careful engineering of a suitable platform.
Strategies for a successful electronic decoupling of an adsorbed molecule from the metal surface have been developed, \textit{e.\,g.}, ultrathin films of oxides \cite{science_299_542,prb_77_205430,prl_105_217402,acsnano_8_54} and salt \cite{nature_531_623,nature_538_623,prl_118_127401,natcommun_8_15225,prl_119_013901}, molecular buffer layers \cite{prl_92_086801,prl_95_196102,natphoton_4_50,oe_17_2714,cpc_11_3412,prb_86_035445,nl_13_2846,natcommun_6_8461,nl_16_2084,acsnano_11_1230}, lifted \cite{jacs_135_15794} and suspended \cite{prl_112_047403,nl_16_6480,prl_116_036802,nl_18_175} molecules.

Somewhat surprisingly, STM-induced fluorescence of isolated single molecules has exclusively been reported for porphyrins \cite{science_299_542,prb_77_205430,prl_105_217402,nl_16_6480} and phthalocyanines \cite{nature_538_623,nature_531_623,prl_118_127401,natcommun_8_580,natcommun_8_15225,prl_119_013901} to date and it remains an open question whether other classes of molecules are suitable for the observation of electroluminescence induced by an STM\@.
An increased number of classes giving rise to single-molecule luminescence is highly desirable to develop a clear picture of the mechanism of light emission at the molecular scale.
Currently, two competing scenarios are being discussed.
The first model \cite{science_299_542,prb_77_205430,prl_105_217402,prl_112_047403,nature_531_623,natcommun_8_580} requires the attachment of electrons and holes to, respectively, unoccupied and occupied states of the molecule by aligning the electrode Fermi levels with the molecular resonances.
The resulting electron-hole pair radiatively recombines and gives rise to the molecular fluorescence.
The second model, in contrast, does not require the alignment of electronic levels.
Rather, it relies on an energy transfer -- possibly mediated by the localized tip-induced plasmon -- from the inelastic tunneling electrons to the molecular emitter for subsequent radiative exciton decay \cite{natphoton_4_50,prl_116_036802,nl_16_6480,prl_118_127401,prb_84_205419}. 
In both models, the tip-induced plasmon enhances the light emission \cite{pr_69_37}.

Here, we report luminescence from single $5$,$10$,$15$,$20$-tetraphenylbisbenz$[5$,$6]$indeno$[1$,$2$,$3$-\textit{cd}:$1'$,$2'$,$3'$-\textit{lm}$]$perylene (C$_{64}$H$_{36}$, DBP), which is currently attracting attention as an organic electron donor in photovoltaic applications due to its high absorption coefficient \cite{apl_99_153302,apl_102_073302,apl_102_143304,pccp_16_8852} and as an assistant dopant in organic light-emitting diodes for its high fluorescence quantum yield \cite{semsc_93_1029,apl_102_073302,natcommun_5_4016}. 
In addition -- as we demonstrate in this work -- it represents a new class of molecules for STM-LE\@.
Indeed, it belongs to the family of polycyclic aromatic hydrocarbons, which has revealed fluorescence in frozen matrices \cite{jpca_104_1,jpcb_106_910} suggesting its potential suitability in STM-LE\@. 
The most appealing property of DBP is the presence of a single transition dipole moment that is oriented along the long symmetry axis of the molecule.
This helps reducing the complexity of luminescence spectra, which may be a superposition of photon emission from more than a single transition dipole moment in the cases of porphyrins \cite{science_299_542,prb_77_205430,prl_105_217402,nl_16_6480} and phthalocyanines \cite{nature_538_623,nature_531_623,prl_118_127401,natcommun_8_15225,prl_119_013901}.  
The present experiments further unveil the dependence of the photon yield on the intramolecular site as well as the occurrence of vibrational progression.  
More importantly, by varying the lateral tip--molecule distance photon spectra were recorded from at-distance to on-molecule excitation of light.
The spectroscopic line shape of the emitted light exhibits a Fano profile whose detailed shape progressively evolves with the lateral tip--DBP separation.
This observation hints at describing \textit{both} on-molecule \textit{and} at-distance excitation of molecular fluorescence within the Fano picture.

\begin{figure}
\includegraphics[width=0.95\textwidth]{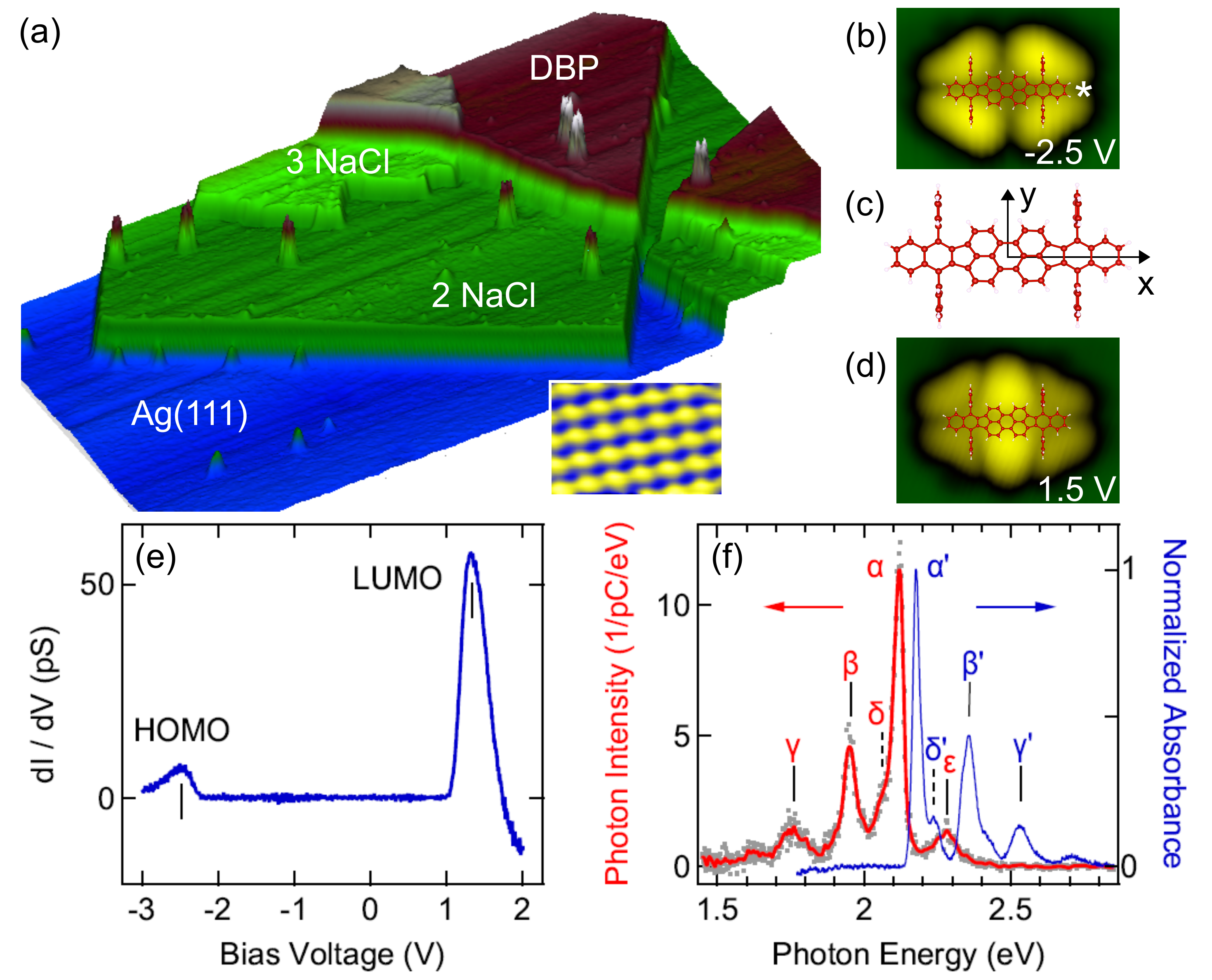}
\caption{ 
(a) Pseudo-three-dimensional representation of an STM image of Ag(111) covered with NaCl islands $2$ and $3$ atomic layers high (bias voltage: $-2.5\,\text{V}$, tunneling current: $5\,\text{pA}$, size: $150\times 75\,\text{nm}^2$)\@.
At $-2.5\,\text{V}$, $2$-layer ($3$-layer) NaCl islands exhibit an apparent height of $\approx 270\,\text{pm}$ ($\approx 385\,\text{pm}$)\@.
Inset: Atomically resolved STM image of a $2$-layer NaCl island ($-2.5\,\text{V}$, $5\,\text{pA}$, $2.4\times 1.7\,\text{nm}^2$)\@.
(b) STM image of single DBP on a $3$-layer NaCl island ($-2.5\,\text{V}$, $5\,\text{pA}$, $3.6\times 2.6\,\text{nm}^2$) with superimposed ball-and-stick model of DBP (to scale)\@.
The asterisk indicates the position where the fluorescence spectrum in (f) was acquired.
(c) Ball-and-stick model of DBP in its relaxed vacuum structure with $x$ and $y$ axes denoting the symmetry axes of the molecule.
C (H) atoms appear red (white)\@.
The distance between the outermost H atoms along $x$ is $2.16\,\text{nm}$.
The molecule has a $D_{2h}$ symmetry with a planar dibenzoperiflanthene backbone.
The phenyl rings are perpendicular to the backbone plane and tilted by $0.5^\circ$ towards the DBP center with respect to the $y$ axis.
The optimized structure was obtained by density functional calculations at the B$3$LYP/$6-311$G$++$(d,p) level \cite{langmuir_32_1981}.
(d) Same as (b), $1.5\,\text{V}$\@.
(e) Constant-height spectrum of $\text{d}I/\text{d}V$ acquired at the DBP center.
The feedback loop was disabled at $-3\,\text{V}$, $20\,\text{pA}$\@.
(f) Left: Single-DBP luminescence spectrum (dots) recorded atop the site indicated in (b) by an asterisk ($-2.5\,\text{V}$, $300\,\text{pA}$, spectrum acquisition time: $60\,\text{s}$)\@.
The red solid line represents smoothed data and serves as a guide to the eye.
Right: Normalized absorbance (blue line) obtained for DBP molecules embedded in an Ar matrix at $12\,\text{K}$ \cite{pccp_17_30404}.
}
\label{fig1}
\end{figure}

Figure \ref{fig1}(a) shows an STM image of NaCl-covered Ag(111) with individual DBP molecules atop $2$-layer and $3$-layer NaCl islands.
Low-temperature deposition of DBP led to separately dispersed monomer molecules on the different NaCl islands.
Occasionally, clusters of DBP with different sizes (dimers, trimers) were imaged.

Depending on the bias voltage polarity, individual DBP molecules appear with different internal structure.
At $-2.5\,\text{V}$ [Fig.\,\ref{fig1}(b)] the molecule appears with four bright protrusions that are separated by two crossing nodal lines.
The bright protrusions exhibit their strongest weight in the vicinity of the phenyl groups. 
The long nodal line coincides with the long symmetry axis of the molecule [$x$ in Fig.\,\ref{fig1}(c)]\@.
The short nodal line [along $y$ in Fig.\,\ref{fig1}(c)] is perpendicular to the long molecular axis and intersects it at the DBP center.
STM images of DBP at $1.5\,\text{V}$ [Fig.\,\ref{fig1}(d)] reveal two additional protrusions at the DBP center.
The nodal line along the $x$ axis is preserved and two additional nodal lines oriented nearly parallel to $y$ separate the central protrusions from the protrusions that are almost centered at the phenyl groups.
Owing to the NaCl film, which acts as a buffer layer, the DBP molecules are well decoupled from the Ag(111) surface.
Therefore, it is likely that the adsorbed DBP molecules adopt the vacuum structure [Fig.\,\ref{fig1}(c)]\@.

In order to identify the orbital origin of the structural motifs visible in the STM images at different bias voltage, $\text{d}I/\text{d}V$ spectra were acquired.
Figure \ref{fig1}(e) shows a spectrum that was recorded atop the DBP center.
Peaks at $\approx -2.47\,\text{V}$ and $\approx 1.33\,\text{V}$ are separated by a gap region of $\approx 3.80\,\text{eV}$ where the $\text{d}I/\text{d}V$ signal vanishes.
We attribute the peaks at negative and positive bias voltage to the signatures of the frontier orbitals, \textit{i.\,e.}, to the highest occupied molecular orbital (HOMO) and the lowest unoccupied molecular orbital (LUMO), respectively.
Consequently, the submolecular patterns visible at negative [Fig.\,\ref{fig1}(b)] and positive [Fig.\,\ref{fig1}(d)] bias voltage are related to the spatial density-of-states (DOS) distribution of the HOMO and LUMO, respectively.
The peak positions of HOMO and LUMO vary by about $0.1\,\text{eV}$ across the molecule, which may be related to spatially varying intramolecular charge distribution that entails different extents of screening \cite{jpcm_20_184001}.
Variations of orbital energies in a single molecule were likewise reported for C$_{60}$ \cite{prb_66_161408,prl_90_096802,prb_70_115418} and endohedral C$_{82}$ \cite{prl_94_136802}.
In addition, HOMO and LUMO signatures appear with different strengths in $\text{d}I/\text{d}V$ spectra acquired at different positions atop the molecule, which reflects the spatial distribution of the frontier orbital DOS within the molecule.

Figure \ref{fig1}(f) shows a fluorescence spectrum acquired at one of the ends of the molecular backbone [asterisk in Fig.\,\ref{fig1}(b)]\@.
Well discriminable peaks are observed at $\approx 2.12\,\text{eV}$ ($\alpha$), $\approx 1.95\,\text{eV}$ ($\beta$) and $\approx 1.77\,\text{eV}$ ($\gamma$)\@.  The principal peak $\alpha$ additionally exhibits a shoulder at $\approx 2.07\,\text{eV}$ ($\delta$)\@.
Except for the peak at $2.28\,\text{eV}$ ($\varepsilon$), the fluorescence peaks $\alpha$ --- $\delta$ correspond to the intramolecular transitions that were previously probed in absorption spectra of DBP isolated in rare-gas (Ar, Ne) matrices at cryogenic temperatures \cite{pccp_17_30404} [$\alpha'$ --- $\delta'$, blue line in Fig.\,\ref{fig1}(f)]\@.
In particular, the main peak $\alpha$ appearing in the luminescence spectra [Fig.\,\ref{fig1}(f)] corresponds to the $S_1(\nu'=0)\longrightarrow S_0(\nu=0)$ transition ($\nu,\nu'=0$ denote the vibrational ground states of $S_0$, $S_1$), which in the absorption spectrum occurs at $2.18\,\text{eV}$ ($\alpha'$)\@.
The difference of $\approx 0.06\,\text{eV}$ may be rationalized in terms of the different dielectric environments probed in the luminescence and optical absorption experiments.
The red-shift of peaks $\beta$, $\gamma$, $\delta$ with respect to $\alpha$ in the fluorescence spectrum essentially matches the blue-shift of the vibronic features $\beta'$ ($2.36\,\text{eV}$), $\gamma'$ ($2.53\,\text{eV}$) and $\delta'$ ($2.24\,\text{eV}$) with respect to $\alpha'$ ($2.18\,\text{eV}$) in the absorbance spectrum.
Therefore, the peak $\beta$ and the shoulder $\delta$ to the principal peak $\alpha$ are due to transitions involving several fundamental vibrations [$S_1(\nu'=0)\longrightarrow S_0(\nu=1)$] closely spaced in energy \cite{pccp_17_30404}.
In addition, following the findings of the aforementioned absorption experiment \cite{pccp_17_30404}, combinations of fundamental vibrations as well as second harmonics [$S_1(\nu'=0)\longrightarrow S_0(\nu=2)$] contribute to $\gamma$.

The energy of the weak fluorescence peak at $2.28\,\text{eV}$ ($\varepsilon$), which was not reported in the absorption experiments \cite{pccp_17_30404}, is compatible with light emission from non-thermalized excitons, a photon emission process that is generally referred to as hot electroluminescence where vibrationally excited states of $S_1$ relax to the vibrational ground state of $S_0$, \textit{i.\,e.}, $S_1(\nu'=1)\longrightarrow S_0(\nu=0)$\@.
The apparent violation of the empirical Kasha rule \cite{dfs_9_14}, which forbids molecular luminescence due to non-thermalized excitons may be rationalized in terms of a strongly reduced fluorescence lifetime.
Conventional chromophores in solution or in the condensed phase exhibit a fluorescence relaxation time on the order of $1\,\text{ns}$, while the vibrational relaxation time in an electronic state can be an order of magnitude smaller \cite{jpca_106_9837,jpca_119_1267}.
The Purcell effect (\textit{vide supra}) entails a strong reduction of the fluorescence lifetime in plasmonic environments.
Similar observations of luminescence peaks blue-shifted from the $S_1(\nu'=0)\longrightarrow S_0(\nu=0)$ transition were reported from multilayers of phthalocyanines \cite{natphoton_4_50} and single-porphyrin molecular devices \cite{nl_16_6480}.
The spectral line shape of the tip-induced plasmon may likewise contribute to $\varepsilon$.
In the Supporting Information, molecular fluorescence spectra are shown for different plasmon line shapes (Fig.\,S1, S2)\@. 

Luminescence and fluorescence spectra presented in this work were acquired at $-2.5\,\text{V}$\@.
We found that bias voltages $\leq -2.09\,\text{V}$ led to fluorescence light with virtually identical quantum yield (Fig.\,S3)\@.
Therefore, resonant tunneling of a hole into the HOMO whose signature peaks at $\approx -2.5\,\text{V}$ [Fig.\,\ref{fig1}(e)] is not required.
For voltages $>-2.09\,\text{V}$ the tunneling junction exhibited instabilities which may be traced to the HOMO--LUMO gap and the concomitantly reduced tip--DBP distance at these voltages.
At positive bias voltage, molecular fluorescence was quenched while plasmonic luminescence was not.
We will tentatively rationalize this observation after introducing the Fano picture of fluorescence (\textit{vide infra})\@.

\begin{figure}
\includegraphics[width=0.55\textwidth]{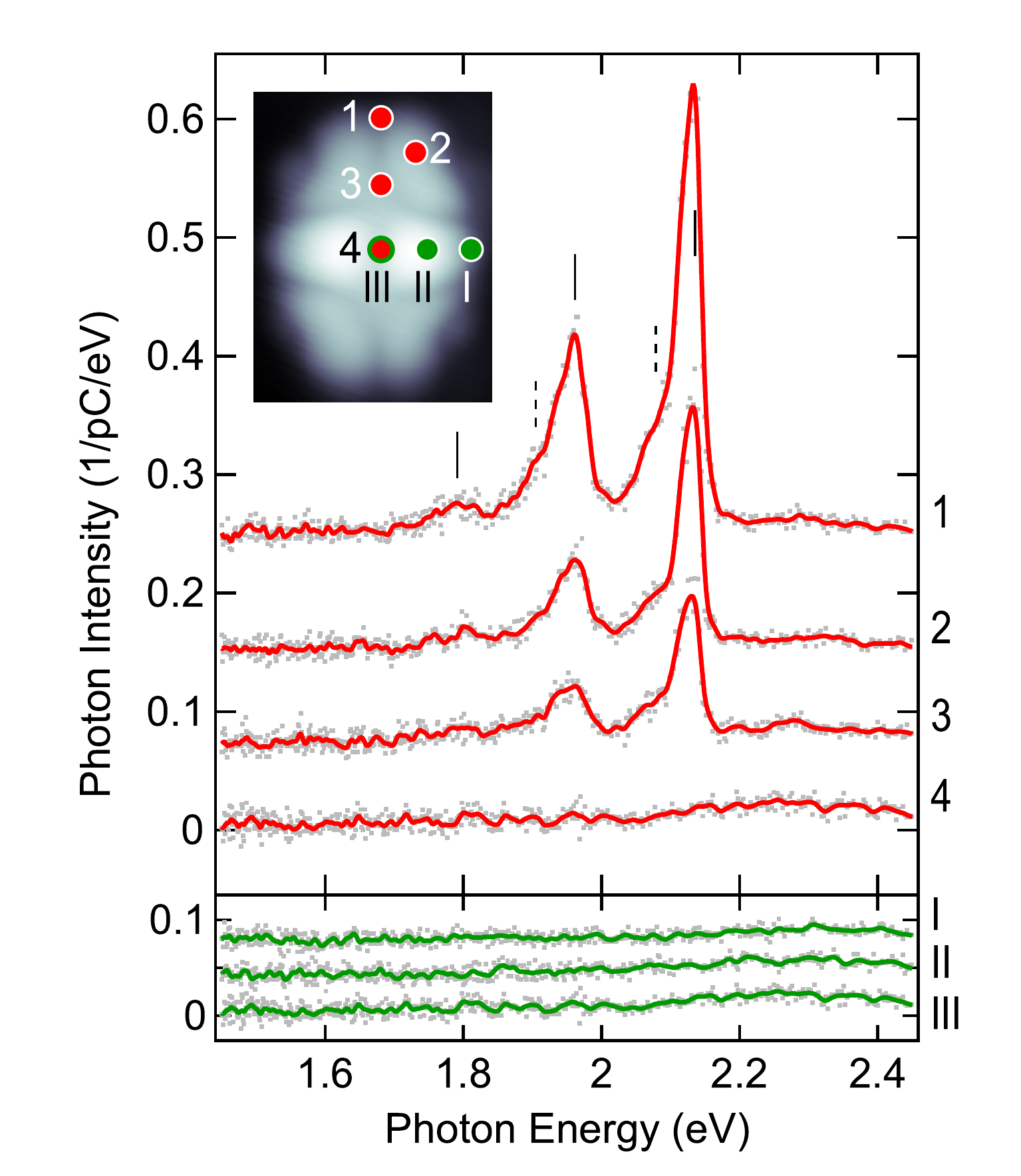}
\caption{
Spatially resolved fluorescence spectra ($-2.5\,\text{V}$, $200\,\text{pA}$, $120\,\text{s}$) at the indicated positions in the STM image.
The main peaks (solid vertical lines) occur at $1.79\,\text{eV}$, $1.96\,\text{eV}$, $2.13\,\text{eV}$\@.
Weak shoulders to the principal transition peak ($\approx 2.08\,\text{eV}$) and its vibrational-progression partner ($\approx 1.91\,\text{eV}$) are indicated by dashed lines.
Inset: STM image of a single DBP molecule on a $3$-layer NaCl island ($1.5\,\text{V}$, $5\,\text{pA}$, $2.6\times 3.3\,\text{nm}^2$) with labels of spectroscopy positions.
}
\label{fig2}
\end{figure}

Spatially resolved DBP luminescence spectra are collected in Fig.\,\ref{fig2}\@.
The strongest emission is observed from the ends of the long molecular axis (position $1$ in the inset to Fig.\,\ref{fig2})\@.
This result may have been expected since the symmetry of the first excited singlet state $S_1$ is $B_{3u}$ while the ground state $S_0$ exhibits $A_g$ symmetry \cite{pccp_17_30404}; that is, the transition dipole moment is oriented along the long axis of DBP and, thus, parallel to the surface.
In accordance with a previously disseminated idea \cite{prl_105_217402,nature_531_623,prl_119_013901,nl_18_2358} the net transition dipole moment of the entire tip--molecule--surface cavity is nonzero at the ends of the molecular backbone, while it vanishes at its center.
Indeed, the luminescence intensity is attenuated towards the center of DBP ($2$ --- $4$) where it becomes too small to be detected.
Possibly emitted light stays below the detection limit, too, along the short axis ({\sf{I}} --- {\sf{III}})\@.

The spectra of Fig.\,\ref{fig2} were acquired with a tip exhibiting a different plasmonic signature than for the spectra of Fig.\,\ref{fig1}(f), which was modified by intentional changes in the microscopic structure of the tip apex \cite{prl_89_156803,prb_79_075406,prb_65_165405}.
The peak at $1.79\,\text{eV}$ [$\gamma$ in Fig.\,\ref{fig1}(f)] is hardly visible and light from non-thermalized excitons [$\varepsilon$ in Fig.\,\ref{fig1}(f)] does not surmount the detection limit in this case.
While the fluorescence spectra depicted in Fig.\,\ref{fig1}(f) and Fig.\,\ref{fig2} represent raw data, the Supporting Information shows the effect of normalizing raw light spectra by the plasmon signature (Fig.\,S1, S2)\@.

\begin{figure}
\includegraphics[width=0.95\textwidth]{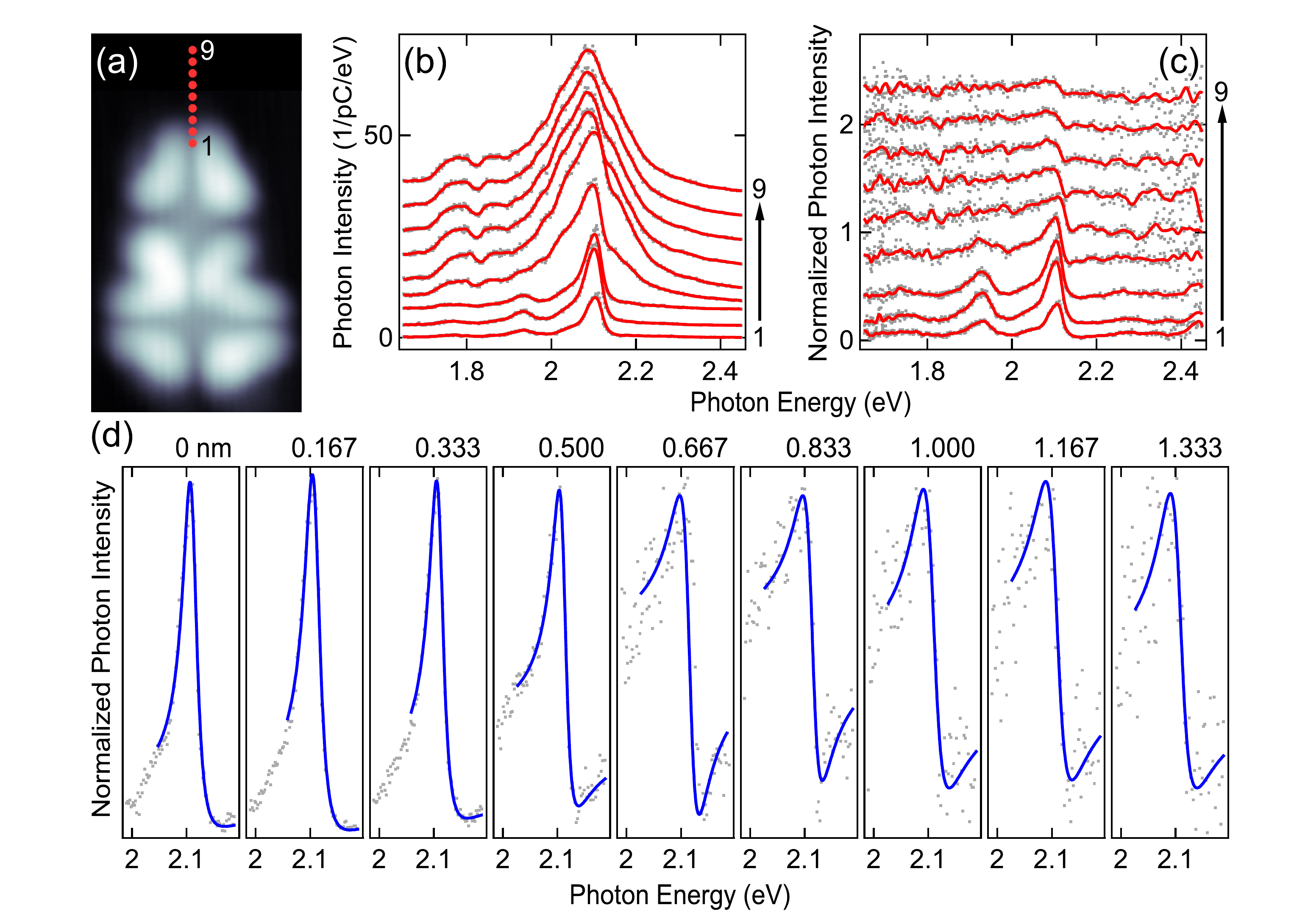}
\caption{ 
(a) STM image of a DBP dimer on a $2$-layer NaCl island ($-2.5\,\text{V}$, $10\,\text{pA}$, $3.3\times 5.6\,\text{nm}^2$)\@.
(b) Luminescence spectra ($-2.5\,\text{V}$, $300\,\text{pA}$, $180\,\text{s}$) recorded along the line defined by the equidistantly ($0.167\,\text{nm}$) arranged red markers in (a)\@.
Distance $d=0\,\text{nm}$ corresponds to the bottom marker in (a)\@.
The spectra are arranged from $d=0\,\text{nm}$ (bottom) to $d=1.333\,\text{nm}$ (top)\@.
Raw data appear as dots.
The solid line represents smoothed data and serves as a guide to the eye.
(c) Luminescence data of (b) normalized by the (smoothed) plasmon line shape.
The solid lines present smoothed data.
(d) Normalized luminescence spectra of (c) plotted on a spectral range constrained to the main $S_1\longrightarrow S_0$ transition at $\approx 2.11\,\text{eV}$\@.
The solid line in each spectrum is a Fano line shape [Eq.\,(\ref{eq1}), (\ref{eq2})] that was fit to the raw normalized data (dots)\@.
Distances are indicated at the top of each panel.
}
\label{fig3}
\end{figure}

The results to be discussed next represent the most important result of this work.
They address the nature of the complex interplay between the inelastic tunneling current, the tip-induced plasmon and the molecular exciton.
To this end, molecular fluorescence was excited by injecting the tunneling current at a finite lateral distance from the molecular emitter. 
For single DBP on NaCl-covered Ag(111) these experiments were challenging due to the mobility of the monomer molecule.
At the elevated bias voltage ($-2.5\,\text{V}$) that is required for the observation of the luminescence signal, monomers of DBP tend to be displaced when the tip is close to the ends of the long molecular axis at tunneling currents exceeding $\approx 150\,\text{pA}$\@.
In order to probe plasmon--exciton interactions with the tip laterally off the DBP, however, elevated currents are required for a decent signal-to-noise ratio.
Fortunately, DBP dimers [Fig.\,\ref{fig3}(a)] are less mobile than monomers and exhibit similar electronic and optical properties as their monomer counterparts.
In particular, the strongest photon emission is observed at the ends of the molecular backbone in both cases.
The Supporting Information contains a comparison of monomer and dimer electronic and optical properties (Fig.\,S4--S6)\@.
Therefore, the luminescence spectra presented in Fig.\,\ref{fig3}(a) were acquired for a DBP dimer.

The light spectra presented in Fig.\,\ref{fig3}(b) were recorded along the $x$ axis [vertically arranged red dots in Fig.\,\ref{fig3}(a)] starting atop a site well inside the molecule (bottom spectrum) and progressively moving outside the molecule.
Inside the molecule and at its outermost boundary (three bottom spectra) the photon spectra are dominated by the $S_1\longrightarrow S_0$ transitions (\textit{vide supra})\@.
The fourth spectrum from the bottom was recorded slightly outside the molecule and is strongly broadened.
The broadening and the overall intensity increase of the luminescence spectra with increasing tip--molecule distance signals the light emission due to the tip-induced plasmon.
At a distance of $1.333\,\text{nm}$ (top spectrum) from the first acquisition site inside the molecule, the spectral line shape of the tip-induced plasmon is nearly recovered
For distances $d\geq 1.5\,\text{nm}$ (not shown) molecular transitions are no longer discernible and the plasmonic luminescence line shape does not change any more.
Intriguingly, in a photon energy range extending from $2.11\,\text{eV}$ to $2.14\,\text{eV}$ where the intensity of the $S_1(\nu'=0)\longrightarrow S_0(\nu=0)$ transition drops from its maximum to essentially zero the plasmon spectral line shape is distorted even several tenths of a nanometer outside the molecule.   
These findings are compatible with results obtained for phthalocyanine molecules \cite{natcommun_8_15225,prl_119_013901}.
In the Supporting Information the evolution of luminescence spectra along the $y$ axis are presented (Fig.\,S7)\@.

To see the distance-dependent line shape variations of the $S_1(\nu'=0)\longrightarrow S_0(\nu=0)$ transition peak along the $x$ axis more clearly the influence of the tip-induced plasmon is reduced by dividing the luminescence data by the (smoothed) plasmon line shape.
These normalized data are plotted in Fig.\,\ref{fig3}(c) on the same photon energy interval as Fig.\,\ref{fig3}(b)\@.
Besides a strong intensity reduction of the molecular fluorescence the peak due to the $S_1(\nu'=0)\longrightarrow S_0(\nu=0)$ transition is discernible even well apart from the molecular emitter.
Moreover, it exhibits a characteristic line shape, which is best seen on a smaller energy range.
Figure \ref{fig3}(d) displays the evolution of the principal fluorescence peak ranging from $1.99\,\text{eV}$ to $2.19\,\text{eV}$ for distances $d=0\,\text{nm}$ (first acquisition site inside the molecule) up to $d=1.333\,\text{nm}$ outside the molecule.
Obviously, the line shape of the principal peak at $2.11\,\text{eV}$ undergoes characteristic variations upon changing the distance between the emitter and the excitation source.
All distance-dependent line shapes were reproduced by fitting 
\begin{equation}
F(\varepsilon)=a\cdot f(\varepsilon)+b
\label{eq1}
\end{equation}
($a$: amplitude; $b$: horizontal background; $\varepsilon$: photon energy) with the Fano function \cite{nc_12_154} 
\begin{equation}
f(q,\Gamma,\varepsilon_0;\varepsilon)=\frac{\left(q+\frac{\varepsilon-\varepsilon_0}{\Gamma}\right)^2}{1+\left(\frac{\varepsilon-\varepsilon_0}{\Gamma}\right)^2}
\label{eq2}
\end{equation}
[$\varepsilon_0$ ($\Gamma$): energy (full width at half maxium) of the $S_1(\nu'=0)\longrightarrow S_0(\nu=0)$ transition peak; $q$: asymmetry parameter of the Fano theory \cite{nc_12_154}] to the experimental data.
A more sophisticated parameterization of the Fano function was not required owing to the sufficiently asymmetric spectral line shape \cite{rsi_87_103901}. 
The graph of $F$ appears as a solid line in each spectrum of Fig.\,\ref{fig3}(d)\@.

\begin{figure}
\includegraphics[width=0.95\textwidth]{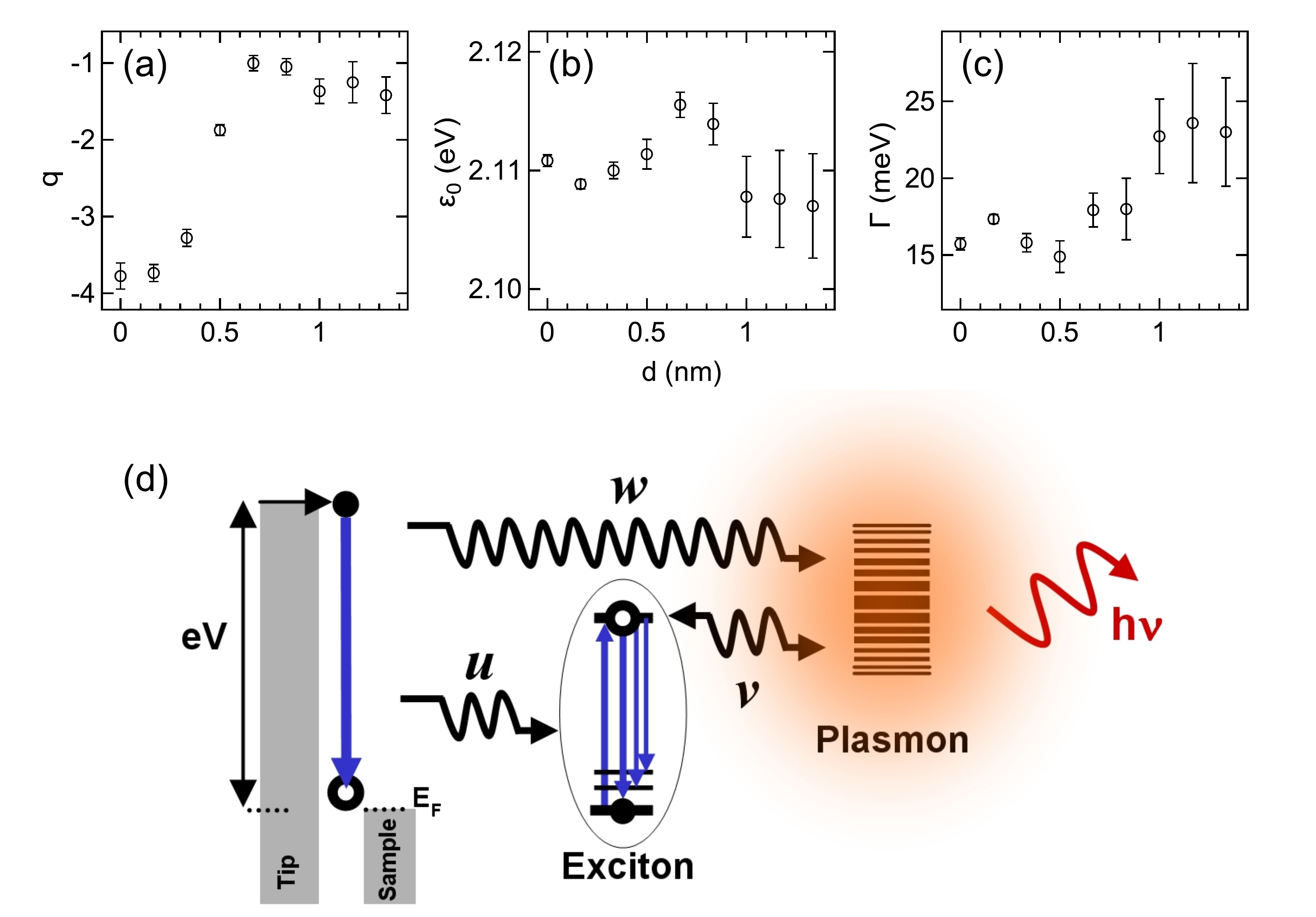}
\caption{ 
(a) Variation of the asymmetry parameter $q$ [Eq.\,(\ref{eq2})] as a function of the lateral distance $d$ between DBP and the tip.
$d=0\,\text{nm}$ is defined by the first acquisition site inside the molecule [bottom red marker in Fig.\,\ref{fig3}(a)]\@.
(b) and (c) Variation of $\varepsilon_0$ and $\Gamma$ [Eq.\,(\ref{eq2})], respectively, with $d$.
(d) Illustration \cite{cnrs_1988} of the coupling mechanism that leads to the Fano line shape in the luminescence spectra of the DBP $S_1(\nu'=0)\longrightarrow S_0(\nu=0)$ transition.
The inelastic tunneling electron couples to the plasmonic continuum of the cavity with a strength $w$ and to the molecular exciton with a strength $u$.
The plasmon--exciton coupling is given by strength $v$.
}
\label{fig4}
\end{figure}

The variations of $q$ are shown in Fig.\,\ref{fig4}(a)\@.
Inside the DBP molecule, at distances $0\,\text{nm}$ and $0.167\,\text{nm}$, $q$ stays at a constant value of $\approx -3.7$\@.
Upon increasing the distance, $q$ increases strongly and levels off at an average value of $\approx -1.2$ for distances larger than $\approx 0.6\,\text{nm}$.
The parameters $\varepsilon_0$ [Fig.\,\ref{fig4}(b)] and $\Gamma$ [Fig.\,\ref{fig4}(c)] stay essentially invariant.
While the actual values of $q$ differ for different tips and, thus, tip-induced plasmons, the trend shown in Fig.\,\ref{fig4}(a) is reproduced.
The Supporting Information collects data sets obtained with different tips and different dimers (Fig.\,S8, S9)\@.
The sign of $q$ depends on the detuning of plasmonic and molecular light spectra.
In our case, the maximum of plasmonic luminescence is at lower energy than the $S_1\longrightarrow S_0$ transition.
In agreement with a previous report, \cite{natcommun_8_15225} the asymmetry factor therefore adopts a negative value.
   
The use of the Fano picture for \textit{both} on-molecule \textit{and} at-distance excitation of fluorescence represents an original idea of this work and shall be discussed in the following.
To this end the plasmonic environment shall be defined as a continuum of states to which both the inelastic tunneling electron and the molecular exciton may couple with strengths $w$ and $v$, respectively [Fig.\,\ref{fig4}(d)]\@.
In addition, the inelastic tunneling electron is coupled to the exciton with strength $u$; that is, our model is based on the assumption that the molecule is excited by direct --- rather than indirect plasmon-mediated --- energy transfer from the injected electron.
This is the physical situation covered by the Fano theory \cite{nc_12_154} giving rise to the spectral line shape $f$ with the asymmetry parameter $q\propto u/(v\cdot w)$\@.
The changes in the asymmetry parameter are induced by a variation of these coupling strengths.
Far away from DBP the inelastic tunneling electron dominantly couples to the plasmon, and the signature of the molecule results from the plasmon--exciton coupling with strength $v$.
In the simple model proposed here the plasmon is considered as a continuum and, therefore, the normalized spectra should reveal a symmetric dip when the tip is positioned sufficiently far away from the molecule. \cite{cnrs_1988}
However, the luminescence line shape adopts a steplike profile, which results from the detuning of the plasmon resonance with respect to the molecular exciton energy. \cite{natcommun_8_15225}
Decreasing the lateral distance between the tip and DBP leads to a reduction of $w/u$, \textit{i.\,e.}, a trading of the coupling strength from $w$ to $u$, and a concomitant increase of $|q|$, as observed experimentally [Fig.\,\ref{fig4}(a)]\@.
The gradual variation of $q$ and the essential invariance of $\varepsilon_0$, $\Gamma$ with changing $d$ suggests --- besides the gradual increase of $u$ at the expense of w --- that the overall luminescence mechanism is the same both atop and aside the molecule. 
In particular, the plasmon--exciton coupling is always present, independent of the lateral tip--DBP distance.
The relative weight of the coupling strengths changes, which leads to a peak-like line shape atop the molecule and to a step-like profile off the molecule.
In the former case, $|q|$ is large, \textit{i.\,e.}, the electron--exciton coupling dominates the electron--plasmon interaction; in the latter case the situation is reversed.

Before concluding we remind that the molecular fluorescence observed for negative bias voltage is suppressed at positive bias voltage to an extent that prevents it from being detected; that is, bipolar electroluminescence is seemingly absent for DBP on NaCl-covered Ag(111)\@.
In at-distance luminescence spectroscopy the plasmon line shape stays essentially invariant at positive bias voltage and does not exhibit the characteristic Fano profiles observed for negative polarity.
In the Fano picture conveyed above [Fig.\,\ref{fig4}(d)] this observation may be explained by a small value of $v$, \textit{i.\,e.}, the inelastic tunneling current only leads to plasmon excitation and its radiative decay.
Both the electron--exciton and the plasmon--exciton interaction are too weak to effectively induce molecular fluorescence.
Details of the underlying mechanisms have not been understood yet.

In conclusion, STM-induced luminescence has been probed for a polycyclic aromatic hydrocarbon molecule, which represents a new class of molecules suitable for electroluminescence with an STM\@.
The spatially resolved photon yield across the molecule reflects the presence of the uniaxial transition dipole moment.
The on-molecule as well as the at-distance excitation of molecular fluorescence gives rise to a characteristic modification of the plasmon emission, which may be described by the Fano theory.
The proposed model of single-molecule fluorescence is based on an energy transfer from the inelastic tunneling electrons to the molecular exciton and / or the localized plasmon.
It shows that the tip--molecule--surface complex acts as an entity in photon emission with a subtle interplay of electron--exciton, electron--plasmon and plasmon--exciton interactions.

\section*{Experimental method}
Experiments were performed with a low-temperature STM operated at $4.7\,\text{K}$ and in ultrahigh vacuum ($5\cdot 10^{-9}\,\text{Pa}$)\@.
Ag(111) surfaces were prepared by Ar$^+$ ion bombardment and annealing.
High-purity NaCl was deposited from a heated crucible onto clean Ag(111) at room temperature, while the cold ($7\,\text{K}$) surface was exposed to a beam of DBP sublimated from molecular powder.
STM images were acquired at constant current and with the bias voltage applied to the sample.
Spectra of the differential conductance ($\text{d}I/\text{d}V$) were recorded by modulating the bias voltage ($10\,\text{mV}_{\text{pp}}$, $750\,\text{Hz}$) and measuring the current response of the tunneling junction with a lock-in amplifier.
The light collection optics and detector is based on a design described \textit{in extenso} elsewhere \cite{prl_116_036802,rsi_80_123704}. 

\begin{acknowledgement}
The authors thank Virginie Speisser and Michelangelo Romeo for technical assistance. 
Financial support by the Deutsche Forschungsgemeinschaft through Grant No.\,KR 2912/12-1, the Agence Nationale de la Recherche (project SMALL'LED No.\,ANR-14-CE26-0016-01, project Labex NIE No.\,ANR-11-LABX-0058\_NIE), and the Centre International de Recherche aux Frontières de la Chimie is acknowledged.
\end{acknowledgement}

\begin{suppinfo}
The Supporting Information is available free of charge on the ACS Publications website at DOI:
Normalization of raw fluorescence spectroscopy data by the plasmon line shape, comparison of DBP monomer and dimer electronic and optical properties, raw fluorescence line shape variations with lateral tip--emitter distance, fluorescence intensity change with tunneling current 
\end{suppinfo}

%\bibliography{ref}
\providecommand{\latin}[1]{#1}
\makeatletter
\providecommand{\doi}
  {\begingroup\let\do\@makeother\dospecials
  \catcode`\{=1 \catcode`\}=2 \doi@aux}
\providecommand{\doi@aux}[1]{\endgroup\texttt{#1}}
\makeatother
\providecommand*\mcitethebibliography{\thebibliography}
\csname @ifundefined\endcsname{endmcitethebibliography}
  {\let\endmcitethebibliography\endthebibliography}{}

\end{document}